# Towards Experimental Nanosound Using Almost Disjoint Set Theory

Cameron L. Jones

Centre for Mathematical Modelling, School of Mathematical Sciences

Swinburne University of Technology

P.O. Box 218, Hawthorn, Victoria 3122, Australia

Fax: +613 9819 0821, Email: cameron.jones@swin.edu.au

**Abstract:**

Music composition using digital audio sequence editors is increasingly performed in a visual workspace where sound complexes are built from discrete sound objects, called gestures that are arranged in time and space to generate a continuous composition. The visual workspace, common to most industry standard audio loop sequencing software, is premised on the arrangement of gestures defined with geometric shape properties. Here, one aspect of fractal set theory was validated using audio-frequency sets to evaluate self-affine scaling behavior when new sound complexes are built through union and intersection operations on discrete musical gestures. Results showed that intersection of two sets revealed lower complexity compared with the union operator, meaning that the intersection of two sound gestures is an almost disjoint set, and in accord with formal logic. These results are also discussed with reference to fuzzy sets, cellular automata, nanotechnology and self-organization to further explore the link between sequenced notation and complexity.

**Introduction:**

The emerging disciplines of molecular electronics, nanotechnology and quantum information science offers new opportunities for acoustic research and the development of experimental sound synthesis or transduction methods. Increasing miniaturization at the nanoscale suggests that classical information can be entangled and that quantum nonlocality will increase the capacity of such systems to perform useful calculations. This paper considers sound information as a form of geometric interaction, and begins with an overview of classic digital sound interactions. One aim of this paper is to suggest how thinking about sound from the perspective of set theory offers a formal way to discretize information. In turn, experiments with nanomaterials offer exciting opportunities for the development of unusual sound synthesis or translation protocols.

An extensive range of music composition software is available for digital sound arrangements (Fink, 2000; Jones, 1999a). Artists as diverse as Jean Michel Jarre, the Chemical Brothers and Cher use computer-based sequencers in their work (Jones, 1999b; Scott, 2000). Software applications like Pro Tools, Cubase, eJay, Fruity Loops or Sonic Foundry's ACID all provide a digital workspace that emulates traditional analog composition methods (Chadabe, 2000). A general feature of software audio sequencers is the visual environment used for manipulating size, position and other visual properties of sound objects, and for adjusting playback options. In practice, many commercial, shareware or freeware audio editors and sequencers are available (Appendix A). Importantly, a sequencer is a digital device that can record, edit and output MIDI, WAV, MP3 or other file messages in a linear sequence using a track-based format, where separate instruments, voices or samples are located on separate tracks.  This interface is analogous to a multitrack tape recorder that follows a linear time line to process audio control events (Miles-Huber, 1999; McGee, 2000a-b).  This paper only considers computer-based software sequencers similar to those detailed in Wentk (1999).

The sequencing process allows the user to select discrete sound objects; most commonly in WAV or MIDI file format and arrange these in time and space.  Figure 1 highlights the important features of this visual approach using a grid representation with the rows representing the different instruments (or WAV files), while the columns represent time.

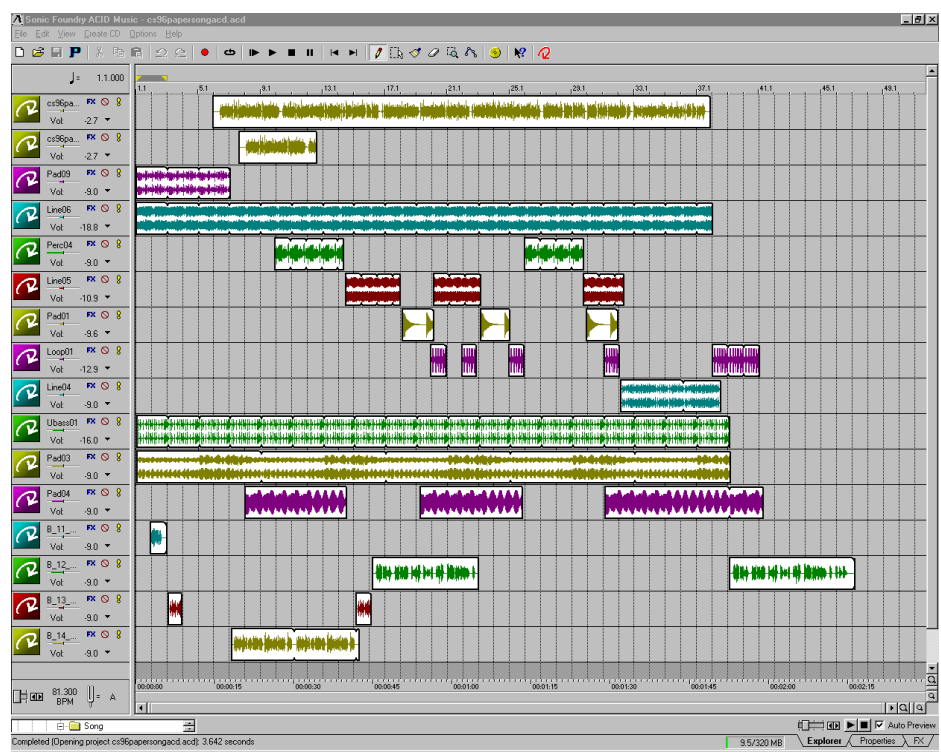

**Figure 1.**  Individual WAV file audio samples are imported, and dragged into their own unique row.  A composition is built by selecting a WAV file and painting it across the screen.  The entire composition is read (played) in parallel from left to right, and may involve the triggering of multiple WAV files at once.  (Screen capture from ACID Music 2.0).

A typical example of this in dance music occurs around a four bar loop where a sequence of sixteen beats is repeated throughout the composition where different instruments, sound fonts or sample files are triggered 'on' or 'off' to alter mood (Jones, 1999b). Layering is another sequencing technique where vocal parts are overlaid on the same track to rapidly repeat a sample (Anderson, 2000). This is a common example of the union operator. Use of multiple track layering naturally involves intersection effects.

Contemporary music composition is becoming increasingly reliant on digital techniques (McGee, 2000a-b), not to mention how 'music' can be mapped to alternative numerical outputs like fractals (Diaz-Jerez, 1999) or cellular automata (Miranda, 1998) or manipulated using granular synthesis processing (Zadorin, 1997; Duesenberg, 1999) to generate diverse effects.

This paper examines the fundamental logical process of 'Union' and 'Intersection' that underpins all computer sequence-based compositions. This is addressed by investigating one aspect of fractal set theory (Hastings and Sugihara, 1993) that naturally follows from the object-oriented feature of audio sequencing software. For simplicity, we begin by examining straightforward arrangements of sound complexes to explore their *formal logic* properties. It is recognized that in practice, set combinations of different sounds can be highly complex. To this end, *fuzzy logic* may better represent subtle effects of union and intersection (Kosko, 1994), although fundamental results detailed here are likely to hold in more fine-grained interpretations.

Additionally, this paper considers the topic of sonic information structure and why sound is a scholarly topic in computer science. O'Donnell and Bisnovatyi (2000) have previously emphasized the importance of 'sound' research with respect to its' use as an information channel of no less significance that visual data. To this end, this paper considers the mathematical nature of discrete sound samples as part of a larger sound matrix. Only by recognizing formal properties of simple interactions is it then possible to explore sound in more complex systems. My motivation in this direction is towards experiments that explore novel interface opportunities at the nanoscale. At this level of organisation, amplification or transduction of spin-state properties might well offer a new class of audio defined as nanosound that is acoustically unusual due to quantum interactions, inter-molecular vibration or chaotic-type effects.

*Research Question:*

If several WAV files are triggered in parallel, and each sound complex is treated as a set of notes in frequency space, then what happens when discrete WAV files are combined (Union) or overlapped (Intersection)? This problem was examined from the perspective of almost-disjoint set theory that had been addressed theoretically by Hastings and Sugihara (1993). Their discussion focussed on how the scaling dimension (Euclidean or fractal) was influenced by simple logical operations, and general results concerning the definition of an almost disjoint set (Weisstein, 1996-) in terms of discrete objects are summarized in Figure 2. Since digital audio can be discretized, one can imagine different mapping procedures between size and frequency at the nanoscale. Therefore digital signals are topologically equivalent to material interactions between nano-units, where complexity emerges from the bottom-up.

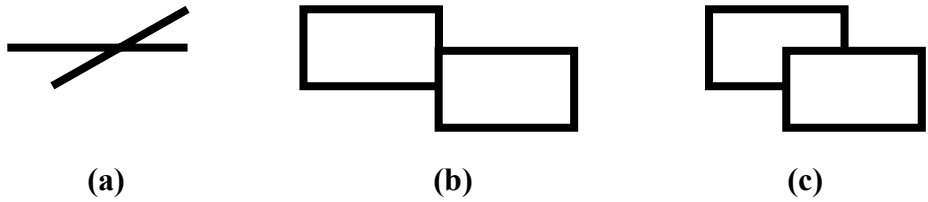

**Figure 2.** **(a)** Two lines form an almost disjoint union and meet at a point. Note that lines have an Euclidean dimension of 'one', d=1, while the intersection point has d=0. Panel **(b)** shows two sound sample objects that are triggered or 'played' from different rows within the sequencing window. The sound font is a plane, and has d=2, while the intersection point is a line with d=1. Panel **(c)** shows the impact of overlap between two sound fonts. The intersection of two planes (d=2), forms another plane (d=2), which is clearly not an almost disjoint union. Re-drawn from Hastings and Sugihara (1993).

*Experiments:*

(i) Perform self-affine analysis to measure the scaling dimension for the UNION set *x* formed by a binary operation on two discrete WAV files, A∪B.

(ii) Perform self-affine analysis to measure the scaling dimension for the INTERSECTION set *x* formed by a binary operation on two discrete WAV files, A∩B.

(iii) Perform visual recurrence analysis on experiments (i) and (ii).

In these experiments I set A=B as the defining property.

*Software:*

Sonic Foundry Acid Music 2.0 was used as the audio sequencer to generate sound complex arrangements that reflected one aspect of fractal set theory.  WAV file data was converted into ASCII text for analysis with Benoit, which performs a range of standard fractal analysis functions.  Nonlinear Dynamics Toolbox (Applied Chaos Lab - Georgia Institute of Technology) was used to perform conversion into numerical (ASCII) output from input WAV files.

*Fractal Analysis:*

Self-affine fractal analysis was performed using Benoit, Version 1.01 (TruSoft Int'l, Inc).  The Variogram, Power-Spectral Analysis and Wavelet Transform methods were used to characterize signal complexity for the union and intersection operator.  For the power law method, the linear trend was removed first from each dataset.  In this context, a self-affine fractal curve is equivalent to plotting audio amplitude versus time.  In all cases, the Hurst exponent was calculated, which is a sensitive indicator of amplitude fluctuation in one-dimensional signals.  The Hurst exponent, H can take values: $H \in (0,1)$.  If $H=\frac{1}{2}$ then the signal sequence is random; if $H>\frac{1}{2}$ then the sequence is persistent, and adjacent amplitude values are likely to be correlated; while if $H<\frac{1}{2}$ then the sequence is antipersistent, and alternate values are likely.  In all cases the substitution $D=2-H$ was used to convert Hurst values into the more common Fractal Dimension, D format.

<u>Variogram Method:</u>

This method is often called the variance of increments method, and evaluates the squared difference between two *y*-values in a graph separated by a distance *w*.

Power-Spectral Analysis:

This method plots the power spectrum versus frequency to estimate scaling at different resolutions.

Wavelet Transform:

The one-dimensional wavelet transform identifies local variations in signal amplitude by decomposition into time-frequency space. This method shares similarity with Power-Spectral analysis, but is not limited to analysis using combinations of trigonometric functions built from sine waves to represent signal properties. In this application, the analyzing wavelet was a step function (Haar wavelet).

*Recurrence Analysis:*

Recurrence plots are a recent qualitative visualization technique (Konov, 1999) that are suitable for reconstructing multidimensional representations from one-dimensional datasets (Eckmann et al., 1987). Highly deterministic signals show well-structured recurrence plots, while random data show more uniform color distributions. Version 4.2 of Visual Recurrence Analysis was used to analyze audio WAV files.

*Audio WAV Set Construction:*

Defining the sound complex trigger sequence for this experiment formed the set x. In this experiment, an arbitrary WAV file was used (Bass02). The UNION and INTERSECTION operations were arranged following Figure 3 below.

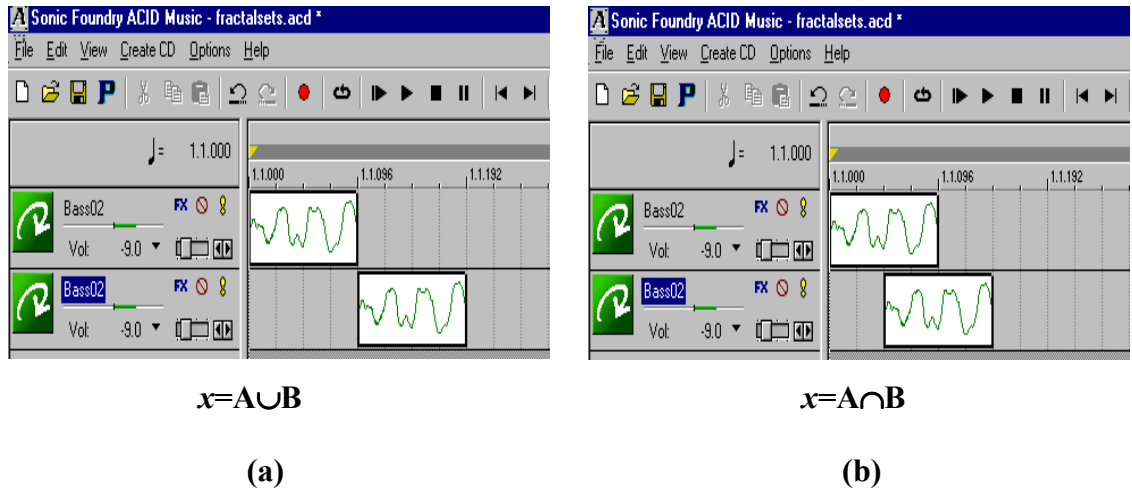

$x=A \cup B$          $x=A \cap B$

(a)          (b)

**Figure 3.** **(a)** The left panel shows the logical union of two WAV files, while the right panel displays **(b)** the logical intersection of the two WAV files. This figure should be examined with reference to Figure 2

Recurrence analysis has been performed for both logical operators to visually demonstrate how the intersection of two sound objects leads to a less complex morphology, compared with the union operator.

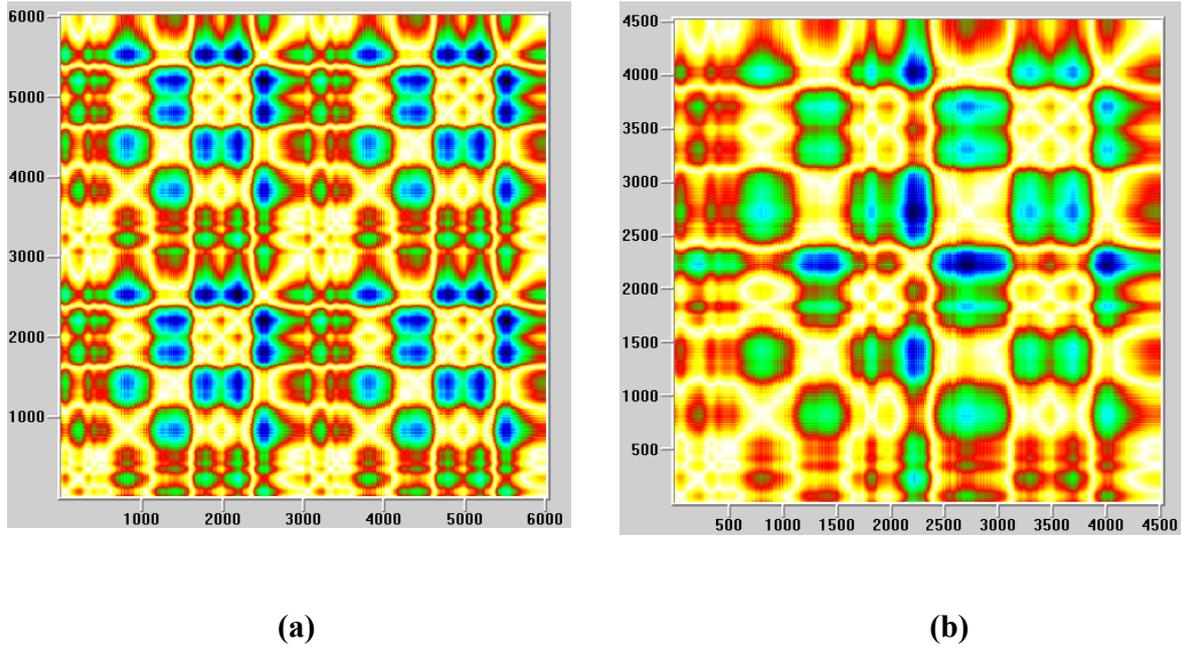

                    **(a)**                                                 **(b)**

**Figure 4.** Recurrence plots of the **(a)** union and **(b)** intersection of the sample WAV files. Delayed coordinate embedding has been used to visualize the spatial correlations of the two sample files. It is clear from the recurrence plots that the intersection operator returns a visual display that is less complex than for the union operator. This visual representation of the data supports results obtained using self-affine scaling methods.

*Definitions:*

Almost Disjoint Sets

A set $x$ is the almost-disjoint union of two sets A and B if $x$ is the UNION of A and B, and the INTERSECTION of A and B has lower dimension than A and B.

Unification of Sets

A new set $x$ generated by the unification of A and B consists of all elements contained in A or in B or in both.

Intersection of Sets

A new set x generated by the intersection of A and B consists of those elements contained in A and in B.

Negation of Sets

A new set x consists of all the elements in some set-valued universe, but not in the set A is the negation of A.

Fuzzy Logic

A multivalued logic allowing intermediate values to be assigned to statements like *yes/no*, *black/white*, or *true/false*. This extends the range of possible observation statements to include descriptions like *closer to*, *more like*, etc.

Binary Operations

A rule assigning to two elements *x,y* of a set, an element x o y of the same set. Addition, multiplication, division and subtraction are binary operations. (Daintith and Nelson, 1989).

Cellular Automata

Computer simulations or models showing homogeneous interactions that occur on a grid or lattice. Each grid point is equivalent to a cell state. Rules are used to specify cell state and cell state evolution is dependent on the state of the cell and a finite number of neighboring cell states. Rules are applied in parallel, yet local neighborhood interactions occur in discrete steps. (Weimar, 1996).

Self-Organization

The appearance of patterned organization between selected parts in the absence of external constraints.  Used here, we mean that although the composer externally selects different arrangements of musical gestures, the audible result (the sound complex) is a function of union, intersection and negation of waveforms at both local and global scales. (See Self-Organizing Systems FAQ under Internet Resources - Appendix A).

**Results:**

The two data sets constructed as shown in Figure 3 revealed the following general results:

(i) Self-affine scaling analysis was a straightforward and sensitive indicator to quantify

logic statements applied in the audio spectrum.

(ii) A set *x* is the almost-disjoint union of two sets A and B if *x* is the union of A and B (H~0.7, **D**~1.30), and the intersection of A and B has lower dimension (H~0.81, **D**~1.19) than the dimensions of A and B.

(iii) Visual recurrence analysis provided qualitative support that the intersection operator returned a less complex morphology pattern compared with the union.

| Logical Operation | Variogram (Hurst exponent and Fractal Dimension) | Power Spectral Analysis (Hurst exponent and Fractal Dimension) | Wavelet Transform (Hurst Exponent and Fractal Dimension) | Mean Values (All 3 methods) |
|---|---|---|---|---|
| A ∪ B | H=0.71 D=1.30 | H=0.62 D=1.38 | H=0.76 D=1.24 | H=0.70 D=1.30 |
| A ∩ B | H=0.77 D=1.23 | H=0.80 D=1.20 | H=0.87 D=1.13 | D=0.81 H=1.19 |

**Table 1.** Table showing the Hurst exponent and Fractal Dimension estimates following evaluation of self-affine scaling using three methods for union and intersection operators applied to the sound complexes detailed in Figure 3. It is recognized that each WAV file dataset was a self-affine curve, and hence the Hurst exponent is the appropriate statistic, however, Hurst values have been transformed into their equivalent fractal dimensions for purposes of clarity following D=2-H.

**Conclusions:**

Overlap of multiple sound gestures to create a 'sound complex' is a fundamental, yet complex feature of the software-based music composition process. Parallel processing of sound complexes via digital wave synthesis has provided an interesting opportunity to verify fundamental principles of set theory and logic applied to digital music.

While advances in conventional computer circuitry have shaped the direction of contemporary electronic music, nanotechnology and molecular electronics offers exciting potential for the design of new musical tools and alternative modes of expression (Cutler et al., 2000; Wilkinson, 2000). Nanocomputation works by exploiting information transfer at atomic resolution scales that are below 1μm. I suggest that the results reported here may be useful in formalizing novel formal methods to create, control and process audio information in quantum and molecular systems. For example, conductive mesoscopic polymers such as polyaniline could be used to create contact microphones or other transducers, while other structures such as carbon nanotubes might be used for the construction of eleborate networks of sub-microscopic wind instruments. Interaction with these structures as 'data' might therefore take the form of microwave, ultrasound, vibration, optical or conductivity effects to amplify intrinsic properties of these new molecular materials. In this context, Figure 2 should be considered to map adjacent or overlapping nano-units that have been defined in some formal way. At this scale, the digital concept of sound fragment interaction (i.e. union/intersection) is expected to breakdown and fall under the influence of quantum nonlocality. Who can imagine what new sounds occur at the nanoscale, or how existing sound synthesis or processing techniques could benefit from unusual combinations of classic and molecular materials?

The results reported here suggest that audio signals adhere to formal logic, as would be expected, since the object-oriented sequencing process was implemented geometrically in a visual workspace - equivalent to the sound matrix. It should be recognized that the 'union' function reflects the stopping and starting of different musical gestures, while the 'intersection' operator reflects the overlap or layering of different musical gestures. From a geometric viewpoint, a sound complex formed by the union operator is a more complex event than sound complexes formed with the intersection operator. This experiment supports our premise that the defining property for compositions arranged with digital audio sequencers might be understood and numerically evaluated in terms of fractal set theory.

The potential for development of unique musical-instrument transducer systems using breakthrough microtechnologies that support binary operations is self-evident. One example cited by Cutler et al. (2000) reports on miniature computers called 'SmartDust' that are smaller than $1mm^3$. At an even smaller scale, nanomaterials that might form such systems could have major impact on developments in audio performance. Such devices could be programmed using cellular automata, fractal or chaotic translations to achieve innovative sequenced sounds.

From the fractal interpretation we can conclude that sequenced audio compositions are a class of cellular automata, with rule definitions set by the aesthetic of the composer. Further, union and intersection operations may be understood in terms of binary notation as applied to quantum physics. Therefore 'union' is equivalent to a rise in local energy maxima (a transition, spin up), while 'intersection' is equivalent to a fall in local energy (relaxation, spin down). Idealization of set-valued logic processing in nano-scale systems may be of fundamental significance in the design of novel frequency manipulation devices. The congruence between the fractality of sound complexes and its interpretation using information theory, set theory, formal logic or fuzzy logic means that sequenced music as information, must result in a self-organizing process over time.

**Acknowledgements:** The author would like to thank Dr. Warren Baker for critical reading of this manuscript.

**Appendix A - Internet Resources:**

Benoit 1.1 (1997). TruSoft Int'l Inc. St. Petersburg, Fl. USA. www.trusoft-international.com

DeYoung, J.A. (2000). Discrete simulated power law data in the time-domain using the Kasdin and Walter (1992) method visualized by the Visual Recurrence Analysis (VRA) method. http://tycho.usno.navy.mil/powervra.html

eJay Sequencing Software. http://www.ejay.com

Electronic Musician Online. http://www.electronicmusician.com/

Fractal Music Lab.

http://www.fractalmusiclab.com/default.asp

Fuzzy Logic Laboratorium Linz - Hagenberg

http://www.flll.uni-linz.ac.at/main.html

See: Introduction to Fuzzy Logic Course

http://www.flll.uni-linz.ac.at/pdw/fuzzy/introduction.html

Nonlinear Dynamics Toolbox. Applied Chaos Lab, School of Physics, Georgia Institute of Technology. http://www.physics.gatech.edu/chaos/research/NDT.html

Self-Organizing Systems (SOS) Frequently Asked Questions Version 2.7 December 2001. For USENET Newsgroup comp.theory.self-org-sys. http://www.calresco.org/sos/sosfaq.htm#1.1

Shareware Music Machine.

http://www.hitsquad.com/smm/cat/AUDIO_EDITORS/

SmartDust: Autonomous Sensing and Communication in a Cubic Millimeter. Robotics and Intelligent Machines Laboratory, University of California at Berkeley.

http://robotics.eecs.berkeley.edu/~pister/SmartDust/

See also: The Virtual Keyboard exploiting SmartDust technology.

http://www-bsac.EECS.Berkeley.EDU/~shollar/fingeracc/fingeracc.html

Sonic Foundry – Acid Music 2.0. http://www.sonicfoundry.com/index.html

Sonic Foundry ACIDplanet.com http://www.acidplanet.com/

Visual Recurrence Analysis Version 4.2.

http://home.netcom.com/~eugenek/download.html